\documentclass[journal,transmag]{IEEEtran}

\usepackage{cite}
\usepackage{graphicx}
\usepackage{amssymb,amsmath}
\usepackage{color}

\DeclareUnicodeCharacter{2212}{-}

\begin{document}
\bstctlcite{IEEEexample:BSTcontrol}

\title{Quantifying the spin-wave asymmetry in single and double rectangular Ni$_{80}$Fe$_{20}$ microstrips by TR-STXM, FMR and micromagnetic simulations}


\author{
	\IEEEauthorblockN{Santa Pile\IEEEauthorrefmark{1}, Andreas Ney\IEEEauthorrefmark{1}, Kilian Lenz\IEEEauthorrefmark{2}, Ryszard Narkowicz\IEEEauthorrefmark{2}, Jürgen Lindner\IEEEauthorrefmark{2}, Sebastian Wintz\IEEEauthorrefmark{3,4}, \\Johannes Förster\IEEEauthorrefmark{3}, Sina Mayr\IEEEauthorrefmark{5,6}, and Markus Weigand\IEEEauthorrefmark{4}}
	\IEEEauthorblockA{\IEEEauthorrefmark{1}Johannes Kepler University Linz, 4040 Linz, Austria}
	\IEEEauthorblockA{\IEEEauthorrefmark{2}Helmholtz-Zentrum Dresden-Rossendorf, Institute of Ion Beam Physics and Materials Research, 01328 Dresden, Germany}
        \IEEEauthorblockA{\IEEEauthorrefmark{3}Max Planck Institute for Intelligent Systems, 70569 Stuttgart, Germany}
        \IEEEauthorblockA{\IEEEauthorrefmark{4}Helmholtz-Zentrum Berlin für Materialien und Energie, 12489 Berlin, Germany}
        \IEEEauthorblockA{\IEEEauthorrefmark{5}Paul Scherrer Institut, 5232 Villigen PSI, Switzerland}
        \IEEEauthorblockA{\IEEEauthorrefmark{6}Laboratory for Mesoscopic Systems, Department of Materials, ETH Zurich, 8093 Zurich, Switzerland}
        % <-this % stops an unwanted space
        
}

\IEEEtitleabstractindextext{%
\begin{abstract}The asymmetry of spin-wave patterns in confined rectangular Ni$_{80}$Fe$_{20}$ microstrips, both in single and double-strip geometries, is quantified. The results of TR-STXM and micromagnetic simulations are compared. For the TR-STXM measurements and the corresponding simulations the excitation was a uniform microwave field with a fixed frequency of 9.43\,GHz, while the external static magnetic field was swept. In the easy axis orientation of the analyzed microstrip, the results show a higher asymmetry for the double microstrip design, indicating an influence of the additional microstrip placed in close proximity to the analyzed one.
\end{abstract}

\begin{IEEEkeywords}
FMR, magnonics, micromagnetics, mumax3, spin waves, spin-wave imaging, spin-wave symmetry, TR-STXM.
\end{IEEEkeywords}}

\maketitle

% no page numbering, or other headers / footers
\pagestyle{empty}
\thispagestyle{empty}

\IEEEpeerreviewmaketitle

\section{Introduction}
\IEEEPARstart{S}{pin-wave} (SW) dynamics research in confined rectangular nano- and microstructures is important for the rapidly growing fields of magnonics and spintronics \cite{132,019,021,018,006}. It was shown that the SW behavior can be affected by different factors, for example temperature \cite{004} or the design of microstructures \cite{002,135}, which in turn can be used as a manipulating mechanism. In the present work, the focus is put on the fundamental understanding of the SW behavior in confined rectangular structures under uniform excitation, depending on the relative positioning of two microstrips \cite{054}.

The development of planar microresonators/microantennas containing a microloop allows for focusing microwave (MW) magnetic fields such that they are mainly uniform in the region where the microstructure is located \cite{025}. At the same time they offer a very high sensitivity for ferromagnetic resonance (FMR) spectroscopy of small structures in comparison to cavity resonators due to an increased filling factor \cite{023}. Time-resolved scanning transmission x-ray microscopy (TR-STXM) \cite{Weigand2022,027} in combination with planar microresonators enables direct, time-dependent imaging of the spatial distribution of the precessing magnetization across nanometer-thin microstrips during FMR excitation in the GHz frequency range with elemental selectivity \cite{038,054}.

\begin{figure}
    \centering
	\includegraphics[width=2.8in]{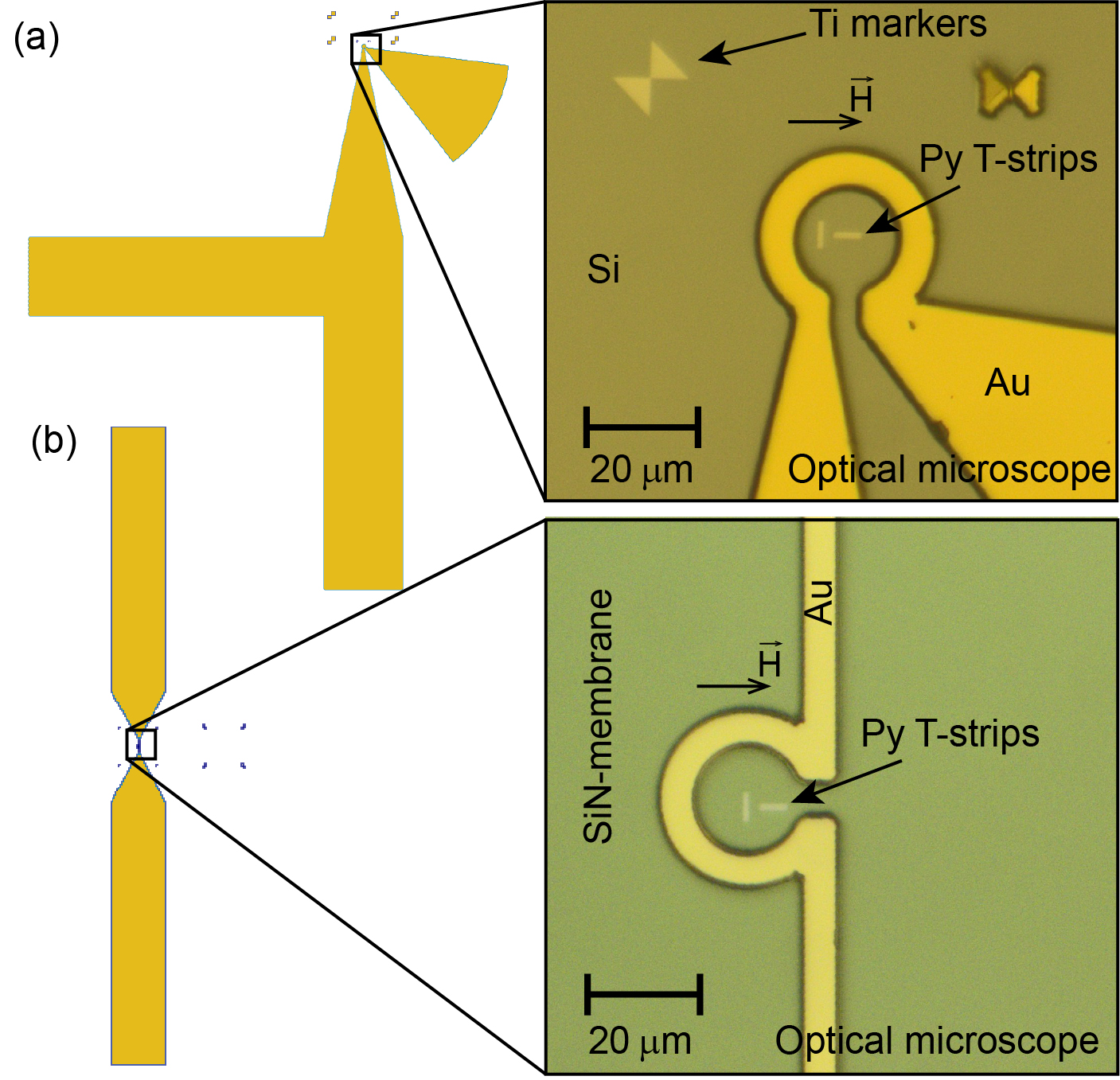}
	\caption{Schematics of the (a) planar microresonator and (b) planar microantenna designs with close-up optical images of the loops with the T-strips samples inside. The direction of the external static magnetic field ($\Vec{H}$) is indicated.}
	\label{antennas}
\end{figure}

\begin{figure}
    \centering
	\includegraphics[width=3.45in]{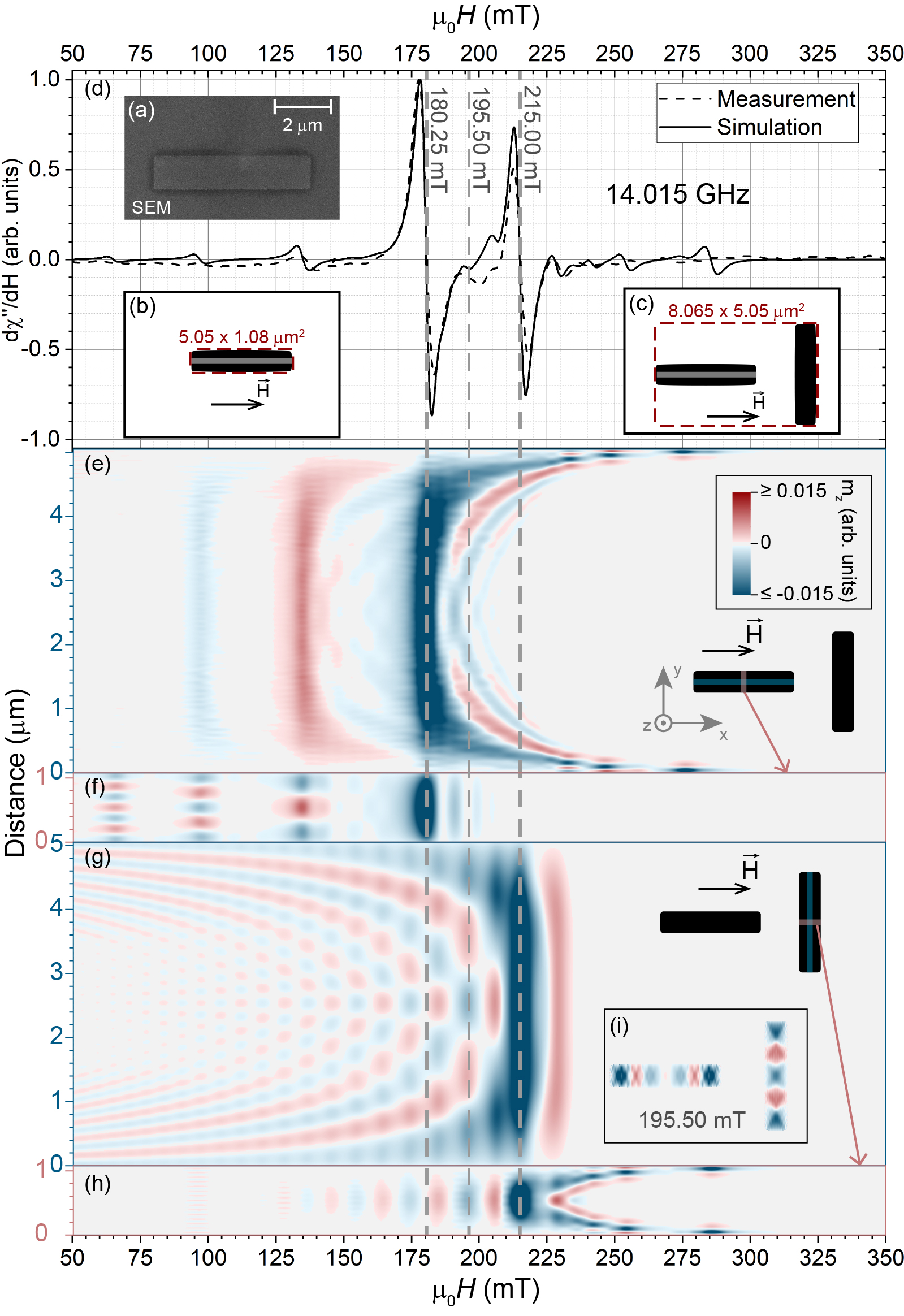}
	\caption{(a) Scanning-electron microscopy image of a single Py microstrip. (b,c)  Single strip and T-strip geometries with their lateral sizes as used for the micromagnetic simulations. The direction of the external static magnetic field is indicated. (d) Simulated (solid) and measured (dashed) FMR spectrum of the Py T-strips at 14.015\,GHz. Overview of the SW profiles (e,g) along and (f,h) across a strip in (e,f) e.a. and (g,h) h.a.\,orientations. (i) An example of the SW patterns at 195.5\,mT.}
	\label{FMRMeasVsSims}
\end{figure}

In general, the confinement of magnetic structures leads to the quantization of SW $k$-vectors along the axis of confinement \cite{020}. The spin-wave spectrum of a uniformly magnetized ellipsoidal magnetic element can be calculated analytically \cite{034a,034b}. However, in most of the cases the magnetic elements used in or considered for applications have a non-ellipsoidal shape. The demagnetizing field and, therefore, the internal magnetic field in rectangular microstrips is strongly inhomogeneous \cite{019}. While it is possible to derive an approximate analytic expression for the general demagnetizing factors of rectangular strips \cite{140}, this does not provide information on the actual spatial distribution of the demagnetizing field in each sample orientation and, additionally, does not take into account edge effects \cite{044}. Micromagnetic simulations, on the other hand, together with spatially resolved imaging can provide access to such kind of information for the investigation of SW dynamics in rectangular microelements \cite{139,038}. In confined structures under uniform excitation, only SW eigenmodes with an odd number of antinodes (amplitude maxima) are expected to occur. This results in a symmetric interference pattern. Changes in the design of the structure, such as presence of an additional rectangular microstrip, can cause change in the internal field configuration and, therefore, symmetry breaking \cite{002, 037}. In this work, the asymmetry quantification of SW dynamics in confined rectangular microstrips by an asymmetry parameter (AP) is suggested and applied to TR-STXM results and micromagnetic simulations.

\section{Experimental details}\label{expDet}
A 30-nm-thick Ni$_{80}$Fe$_{20}$ (permalloy, Py) single strip [Figs.\,\ref{FMRMeasVsSims}(a) and (b)] and double strips (T-strips - see Figs.\,\ref{FMRMeasVsSims}(c) and \ref{antennas}) with a nominal rectangular size of $5\times 1$\,\textmu m\textsuperscript{2} were fabricated on different kinds of substrates depending on the measurements as described in more detail in \cite{054}. The scanning electron microscope (SEM) image of the resulting basic rectangular Py microstrip is shown in Fig.\,\ref{FMRMeasVsSims}(a). The nominal distance between the T-strips is 2\,\textmu m. The designs of the microresonator used for the FMR measurements and the microantenna used for the TR-STXM measurements are shown in Figs.\,\ref{antennas}(a) and \ref{antennas}(b), respectively. Microresonators/microantennas fabrication details can as well be found in \cite{054}.

The microresonator FMR measurements were carried out in a home-built MW spectrometer with field modulation at 78\,kHz using a lock-in technique in field-sweep mode \cite{096,097,130}. In the microresonator, the MW field is oriented perpendicular to the sample plane. The external static magnetic field $\vec{H}$ was applied in the plane of the microstrip. In this geometry, during resonance, the dynamic component of the precessing magnetization is oriented out-of-plane. For the FMR measurements, the frequency was fixed to $f_\mathrm{MW}=14.015$\,GHz, while sweeping the external static magnetic field from -15\,mT to 600\,mT for recording the FMR spectrum.

The TR-STXM experiments were performed at the MAXYMUS endstation of the UE46 undulator beamline at the Helmholtz-Zentrum Berlin during the low-alpha operation mode of the BESSY II synchrotron. For the TR-STXM measurements, the sample is scanned through the focused x-ray beam, while the respective x-ray transmission at each focused point is detected \cite{Weigand2022}. The sample was scanned in steps of 50\,nm. For sensing the dynamic out-of-plane magnetization component $m_\mathrm{z}(t)$ at each scan point, the sample was probed perpendicular to its surface. The sample was scanned in steps of 50\,nm. The photon energy was tuned to the Fe L$_3$-edge ($\sim 708$\,eV). During the TR-STXM measurements a static magnetic field in the range of $65$--$120$\,mT and a small MW field of $\sim$\,$0.5$\,mT were applied along the same axis as for the microresonator FMR measurements. The pump-and-probe measurement scheme allows for probing $m_\mathrm{z}(t)$ at several intermediate points of the precession. The MW frequency of $f_\mathrm{MW}=9.43$\,GHz is phase-locked to the synchrotron frequeny, i.e., the frequency of the x-ray flashes impinging on the sample \cite{031,054}. Hence, TR-STXM images of the $m_\mathrm{z}(t)$ dynamics were taken at 7 points per excitation period.

\section{Results}
\subsection{FMR vs micromagnetic simulations}
In Fig.\,\ref{FMRMeasVsSims}(d), the results of the microresonator FMR measurements at $f_\mathrm{MW}=14.015$\,GHz are shown as dashed black line for the Py T-strips. The direction of the external static magnetic field is indicated by arrows in Fig.\,\ref{FMRMeasVsSims}(c). In the FMR spectrum, one can observe two main resonances with large intensities and several other smaller signals above and below the two main signals. The position of the main FMR resonances and their linewidth were used in order to fit measurement results with the simulated FMR spectra \cite{054}. The resulting simulation parameters for the FMR spectrum shown in  Fig.\,\ref{FMRMeasVsSims}(d) as a solid black line are: cell size of $8 \times 10 \times 7$\,nm$^3$; sample thickness of 27\,nm; no crystalline anisotropy; Py exchange stiffness of 13\,pJ/m \cite{035}; saturation magnetization of 750\,kA/m; Gilbert damping parameter 0.008; static magnetic field ranging from 250 to 0\,mT (see direction in Fig.\,\ref{FMRMeasVsSims}); MW frequency of the uniform out-of-plane field of 14.015\,GHz\cite{038,054} with an amplitude of 0.5\,mT. The sample designs used for the micromagnetic simulations are shown in Figs.\,\ref{FMRMeasVsSims}(b) and \ref{FMRMeasVsSims}(c). The red dashed frames in the figure mark the simulated areas. The lateral sizes are indicated as well.

\begin{figure}
    \centering
	\includegraphics[width=3.2in]{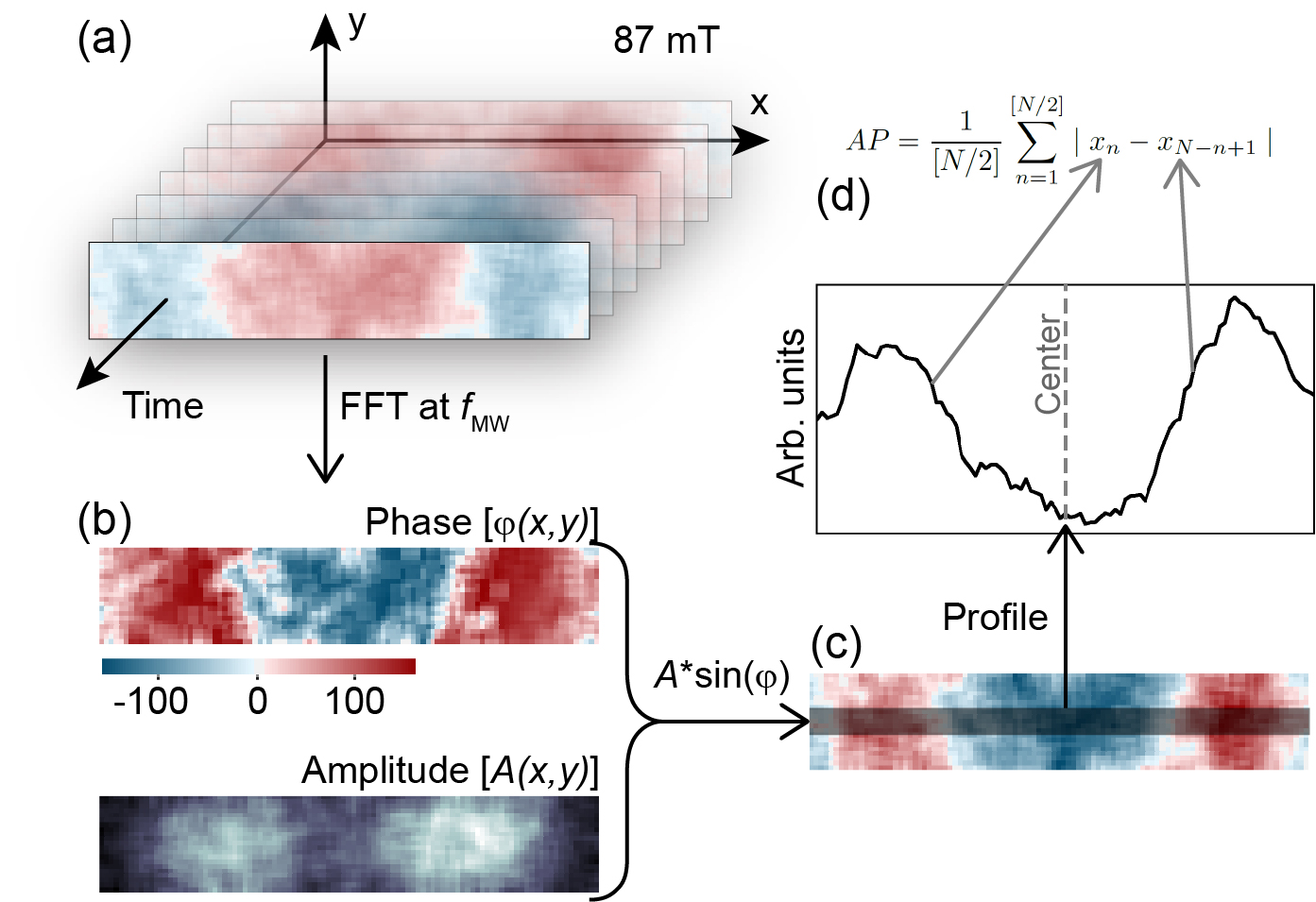}
	\caption{Schematic representation of the data processing cycle up to the AP value calculation. (a) Set of TR-STXM scans at 87\,mT. (b) Spatial distribution of the phase and amplitude obtained from the FFT analysis at $f_{MW}$. (c) SW pattern. (d) SW profile and AP value calculation.}
	\label{APCalc}
\end{figure}

Apart from the FMR spectra, simulations provide an additional information about the spatial distribution of the magnetization. The out-of plane component $m_\mathrm{z}(t)$ reflects the SW dynamics in the microstrips. In Figs.\,\ref{FMRMeasVsSims}(e) to (h) overviews of the simulated $m_\mathrm{z}(t)$ profiles are plotted using the same principle as described in \cite{054} are shown. The overviews display SW profiles along the length [Figs.\,\ref{FMRMeasVsSims}(e) and (g)] and the width [Figs.\,\ref{FMRMeasVsSims}(f) and (h)] of each of the T-strips, in easy axis (e.a., external field parallel to the longer edge of the strip) [Figs.\,\ref{FMRMeasVsSims}(e) and (f)] and in hard axis (h.a., external field parallel to the shorter edge of the strip) [Figs.\,\ref{FMRMeasVsSims}(g) and (h)] orientation, over a range of external field values in correlation with the FMR spectra in Fig.\,\ref{FMRMeasVsSims}(d). According to the micromagnetic simulations, the main resonance lines at 180.5\,mT and 215.3\,mT in Fig.\,\ref{FMRMeasVsSims}(d) correspond to the quasi-uniform FMR excitations. The quasi-uniform mode is the one with almost all magnetic moments across the strip area precessing in phase with the same opening angle. The reason for the nonuniform $m_\mathrm{z}(t)$ spatial distribution closer to the edges \cite{054} is the inhomogeneity of the effective field within the strip. The less pronounced FMR lines correspond to SW excitations with varying amount of amplitude maxima in their interference pattern \cite{037,002,054}. An example of a SW pattern at 195.5\,mT is shown in Fig.\,\ref{FMRMeasVsSims}(i), the corresponding SW profiles are marked with the vertical gray line across Figs.\,\ref{FMRMeasVsSims}(d) to (h). The differences in the resonance fields between the measurement and the simulations at higher fields above 230\,mT can possibly stem from the quality of the edges, i.e.\, the presence of defects etc., which shift the resonance fields of the localized modes to lower values \cite{044,054}.

For the analysis of the TR-STXM results all parameters of the micromagnetic simulations were kept the same, excluding the MW frequency, which was set to $f_\mathrm{MW}=9.43$\,GHz as used in the measurements. At each field value the system was excited for 50 MW periods, first 49 of which were skipped as settling time, and for the last period the spatial distribution of the magnetization is saved in 14 equidistant time frames for further analysis of the spatial maps of $m_\mathrm{z}(t)$, which included fast Fourier transform (FFT) analysis similar to that performed on the measured data.

\subsection{TR-STXM and data analysis}
As described in Section\,\ref{expDet}, the TR-STXM results consist of 7 phase images at each static magnetic field value. Each of it contains the spatially distributed counted x-ray photon signal corresponding to a certain excitation phase over one MW excitation period and, thus, the magnetization precession cycle \cite{038,028}. Each scan includes the Py microstrip and a part of the membrane for the background correction. In general, the raw data includes a static component corresponding to the chemical contrast of the scanned area and the dynamic component corresponding to the magnetic contrast. For each scanned point in space the magnetic part was extracted by dividing the counted x-ray photon signal at each time point by the time averaged value over all time points \cite{030,031,038,073}. Further, a background correction was performed as described in \cite{033}. Upon that, the background part of each scan was removed to proceed only with the data from the Py microstructure. Eventually, the data was filtered in two steps in order to reduce noise. The first step was a FFT filtering at each point in space by converting the signal from the time to frequency domain, filtering out all frequencies except for $f_\mathrm{MW}$, and converting it back to the time domain via inverse FFT. In the second step, the data was filtered in space by replacing every value by the mean value in its range-2 neighborhood.

For both, the simulated and processed TR-STXM data, a temporal FFT at each point of the spatial distribution of the magnetization was performed to extract the spatial amplitude and phase distribution at the given MW frequency \cite{028,129} as depicted in Figs.\,\ref{APCalc}(a) and \ref{APCalc}(b). The extracted amplitude and the phase data was combined into a spatial eigenmodes interference pattern \cite{037} by multiplying the amplitude data with the sine of the phase data as shown in Figs.\,\ref{APCalc}(b) and (c). From the resulting data the central interference pattern profiles ($m_\mathrm{z}(t)$ profiles) were calculated by averaging the central region of the data as shown in Figs.\,\ref{APCalc}(c) and (d). The overviews of the profiles over the range of external static magnetic field were used further to correct the field offset between the simulations and measurements as described in \cite{054}.

\begin{figure}
    \centering
	\includegraphics[width=3.4in]{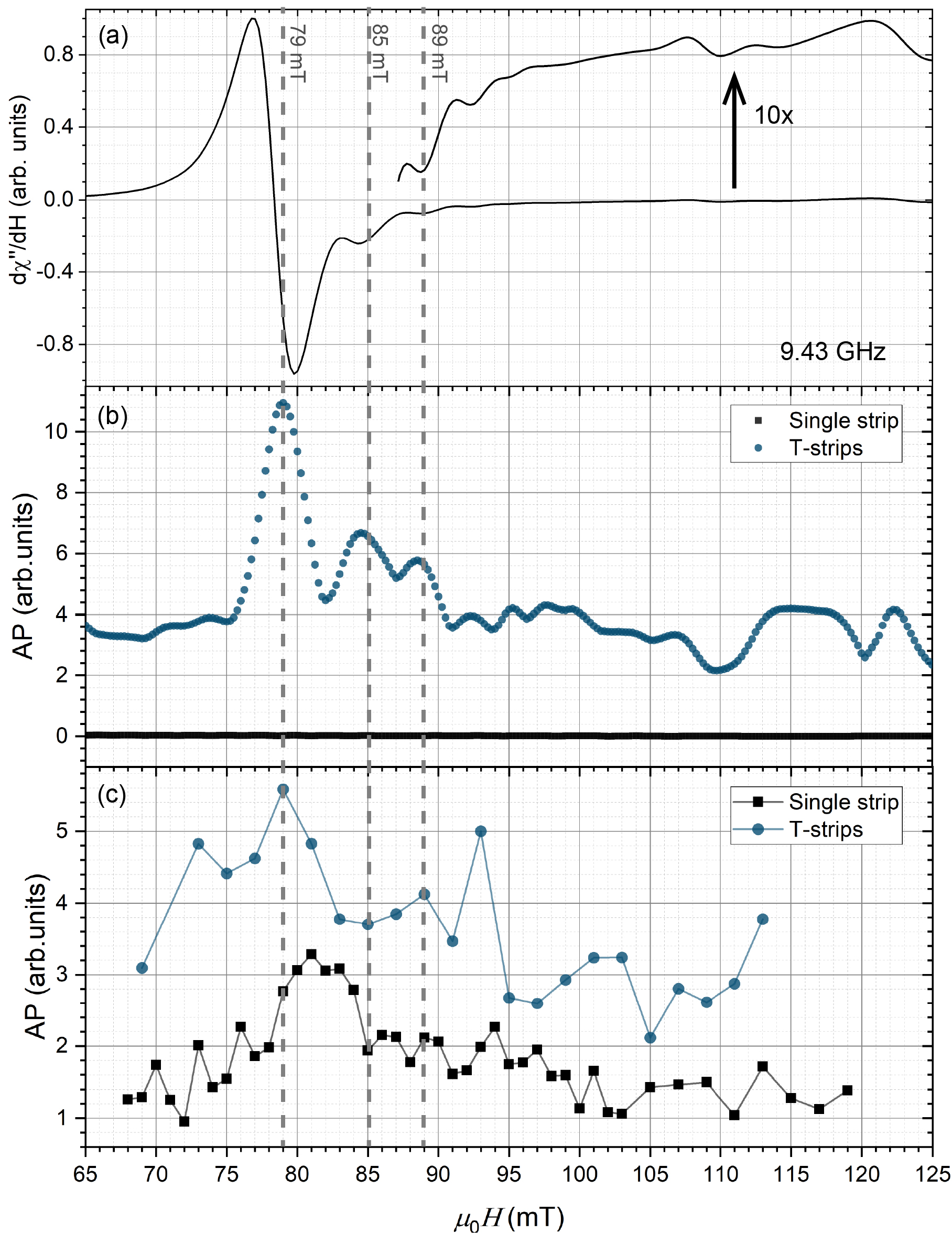}
	\caption{(a) Simulated FMR spectrum of the Py single strip in e.a.\ orientation [see Fig.\,\ref{FMRMeasVsSims}(b)]. (b,c) AP field dependence calculated from the simulated and measured data, respectively. For strips in e.a.\ orientation, the corresponding regions of the strips are marked in Figs.\,\ref{FMRMeasVsSims}(a) and (b), respectively.}
	\label{symmetrySVsT}
\end{figure}

In order to quantify symmetry breaking of the SW patterns an asymmetry parameter (AP) is introduced. It indicates a deviation of the SW pattern from the mirror-symmetric state by analyzing its profile (see Fig.\,\ref{APCalc}). A mirror-symmetric profile here means that the profile is invariant under a reflection about the line in its center (axis of symmetry). The regions of the strips used to calculate the profiles are indicated in Figs.\,\ref{symmetrySVsT}(a) and (b) and Fig.\,\ref{APCalc}(c) with transparent gray rectangles. The AP for a profile consisting of normalized data values $\left\{x_n\right\}_{n=1}^N$ is calculated by
\begin{equation}
	AP = \frac{C}{[N/2]}\sum_{n=1}^{[N/2]} \mid x_{n} - x_{N-n+1}\mid,
	\label{AP}
\end{equation}
where one half of the profile is subtracted from its other half point by point [see Fig.\,\ref{APCalc}(d)]. Here $C$ is a scale factor, when comparing different data, the same value is used. Then the mean value of the absolute values of all differences is taken. Hence, a symmetric profile would give $AP = 0$.

In Fig.\,\ref{symmetrySVsT}(a) a simulated FMR spectrum of the single strip sample in e.a. orientation \cite{054} is shown in order to correlate the calculated AP values with the FMR positions. The calculated AP values from the simulated and measured data are shown in Figs.\,\ref{symmetrySVsT}(b) and (c) for the single strip and the T-strips samples in e.a. orientation, respectively. Overall the $m_\mathrm{z}(t)$ profiles of the single strip sample appear to be more symmetric compared to the T-strips. This indicates that the SWs in one of the T-strips are affected by the other. Nevertheless, some asymmetry is observed in the single strip sample in the measured data as well. The reason could be a small tilt of the strip in the scan or sample defects. The first maximum value of the AP is observed at 79\,mT, close to the quasi-uniform FMR signal. The second and the third maxima are at 84.5\,mt and 88.5\,mT, respectively. They are close to the FMR lines corresponding to the SWs with 3 and 5 amplitude maxima \cite{054}. 

\begin{figure}
    \centering
	\includegraphics[width=3.2in]{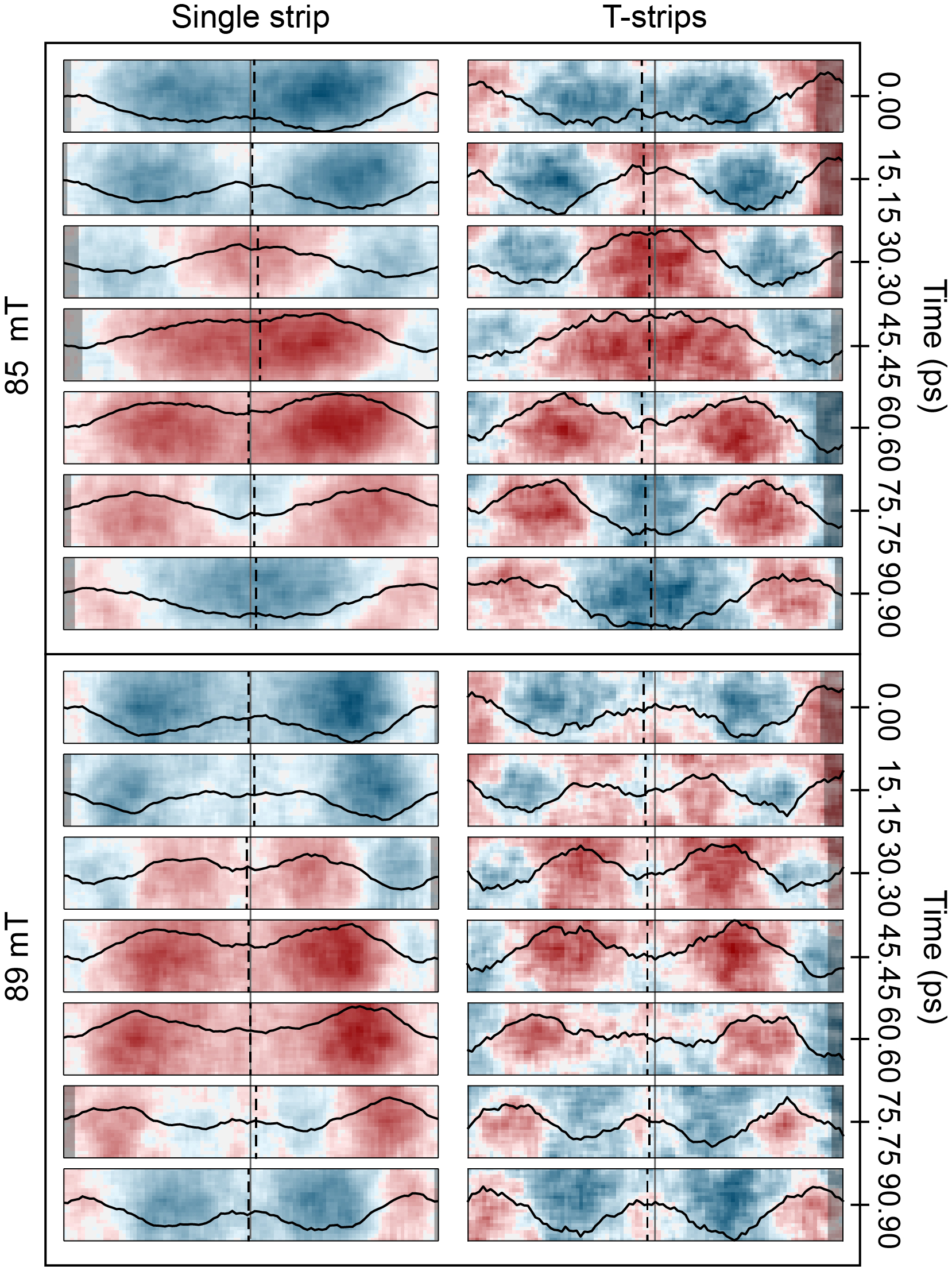}
	\caption{Measured TR-STXM scans of the spin-wave dynamics at 85\,mT and 89\,mT in a strip in e.a. orientation of 2 sample types: single strip and T-strips. Vertical gray solid lines indicate the physical centers of the strips while black dashed shorter lines indicate axis of symmetry for each frame. Transparent rectangles corresponding edge shift (see text).}
	\label{measDynamics}
\end{figure}

In Fig.\,\ref{measDynamics} the spatial distribution maps of $m_\mathrm{z}(t)$ from the TR-STXM scans at 85\,mT and 89\,mT of the single strip and T-strip samples in e.a.\ orientation are shown. Gray vertical solid lines indicate the physical center of a strip. Additionally to the AP value for each phase image shown in Fig.\,\ref{measDynamics}, a axis of symmetry was localized, by finding the edge shift of the data array in Eq.\,\ref{AP}, either left or right, needed to minimize the AP value. The calculated axes of symmetry are indicated with black dashed vertical lines in the figure and the corresponding edge shifts are indicated with transparent rectangles. When looking at the SW dynamics at both fields, one can see that the $m_\mathrm{z}(t)$ patterns are more symmetric with respect to their center for the single strip than for the T-strips. This can be seen even more clearly from the deviation between the physical center and the axes of symmetry in a particular strip. In case of the single strip the axis of symmetry deviates left and right and the absolute value of deviation ranges from 0 to 0.1\,$\mu$m. For the T-strips, the axis of symmetry is always shifted left (the opposite side from where the second strip is located [Fig.\,\ref{FMRMeasVsSims}(c)]) with the absolute values of the deviation ranging from 0.05 to 0.2\,$\mu$m. The reason for that might be inhomogeneous external static and\slash or dynamic magnetic stray fields generated by the second strip.

\section{Conclusion}
In summary, we gained insight on the SW dynamics in confined rectangular Py microstrips, both single and T-strips, using microresonator FMR measurements, TR-STXM imaging with high spacial and temporal resolution with the support of micromagnetic simulations. In order to evaluate the sample design influence on the SW behavior, the asymmetries of the SW patterns were analyzed by means of the introduced AP value. The AP value allows for a sensitive quantification of the symmetry breaking of the SW profiles and, therefore, patterns. The results of the measurements and the simulations show a higher SW asymmetry in the T-strips, in particular for the analyzed strip being in e.a. orientation, when compared to a single strip of the same shape and size. This is an indicator of an influence of one strip onto the SW pattern in the other. The reason can be either an inefficient dynamic coupling between the strips \cite{138} or\slash and the mutual static stray field that changes the effective field distribution within the strips.

\section*{Acknowledgment}
We thank the Helmholtz-Zentrum Berlin for the allocation of synchrotron radiation beamtime. We also thank T.\ Feggeler and H.\ Stoll for their help during the TR-STXM measurements and A.\ Halilovic for valuable contributions to the lithography process. Additionally, we would like to thank M.\ Bechtel for technical support at the beamline. The authors would like to acknowledge financial support from the Austrian Science Fund (FWF), Project No. ESP 4 ESPRIT-Programm and project No. I-3050 during the early stage of the work. S.M. would like to acknowledge funding from the Swiss National Science Foundation (SNSF) under Grant No. 172517.

\bibliographystyle{IEEEtran}
\bibliography{IEEEabrv,refs}
\end{document}